\documentclass{iaus}
\usepackage{graphicx}

\title[Primordial Chemistry] 
{Interstellar Chemistry: Radiation, Dust and Metals}

\author[Marco Spaans]   
{Marco Spaans$^1$}

\affiliation{$^1$Kapteyn Astronomical Institute, University of Groningen, \\ P.O.\ Box 800, 9700 AV, Groningen, the Netherlands \\ email: {\tt spaans@astro.rug.nl} }
\pubyear{2008}
\volume{255}  
\pagerange{119--126}
\jname{Low-Metallicity Star Formation: From the First Stars to Dwarf Galaxies}
\editors{L.K. Hunt, S. Madden \& R. Schneider, eds.}
\begin{document}

\maketitle

\begin{abstract}
An overview is given of the chemical processes that occur in primordial systems
under the influence of radiation, metal abundances and dust surface reactions.
It is found that radiative feedback effects differ for UV and X-ray photons at
any metallicity, with molecules surviving quite well under irradiation by
X-rays.
Starburst and AGN will therefore enjoy quite different cooling abilities for
their dense molecular gas. The presence of a cool molecular phase is strongly
dependent on metallicity. Strong irradiation by cosmic rays ($>200\times$ the
Milky Way value) forces a large fraction of the CO gas into neutral carbon.
Dust is important for H$_2$ and HD formation, already at metallicities of
$10^{-4}-10^{-3}$ solar, for electron abundances below $10^{-3}$.

\keywords{astrochemistry, ISM: molecules, dust, early universe}
\end{abstract}

\firstsection 
\section{Introduction}


In the study of primordial chemistry, and subsequently the formation of the
first stars, it is crucial to understand the ability of interstellar gas to
cool through atomic and molecular emissions and to collapse. Furthermore,
atomic and molecular species allow for probes of the ambient conditions, like
density and temperature, under which stars form. The basic questions of this
contribution are: What sets the abundances of atoms and molecules that cool
gas? What is the role of radiation, metallicity and dust in molecule formation?

A number of different chemical processes are relevant to this effect:

\smallskip
Ion-molecule reactions: A$^+$ + BC$\rightarrow$AB$^+$ + C;

Neutral-neutral reactions: A + BC$\rightarrow$AB + C;

Dissociative recombination: AB$^+$ + e$^-$$\rightarrow$A + B;

Radiative recombination: A$^+$ + e$^-$$\rightarrow$A + $h\nu$;

Radiative association: A + B$\rightarrow$AB + $h\nu$;

Ionization: A + CR/UV/X-ray$\rightarrow$A$^+$ + e$^-$;

Dissociation: AB + UV$\rightarrow$A + B;

Charge transfer: A$^+$ + B$\rightarrow$A + B+;

Grain surface reactions: Grain + A + B$\rightarrow$Grain + AB.
\smallskip

In any chemical network, the above reactions play an important role. For
example, the charge transfer between H$^+$ and O, followed by reactions with
H$_2$ to H$_3$O$^+$, and dissociative recombination with e$^-$, leads to
species like OH and H$_2$O (following certain branching ratios). Similarly
complex routes exist for CO.
In any case, many species are typically joined through different chemical
routes. Thus, it is not trivial to construct concise chemical networks if one
wants to include important molecules like CO and H$_2$O\footnote{It is important to realize that water can be quite an important heating agent in the presence of a warm infrared background, like $T>50$ K dust or a $z>15$ CMB (Spaans \& Silk 2000).}. Of course, in the limit
of low metallicity, chemistry simplifies. Basically, no metals implies no
molecules except for H$_2$ and HD (and a few minor species). Still, even small
amounts of metals and dust ($\sim 10^{-4}$ solar) can be crucial to the
efficient formation of species like H$_2$, HD, CO, H$_2$O and many others,
which is the purpose of this contribution.

\section{Radiation}

The impact of radiation is denoted by UV and X-ray dominated regions
(PDRs and XDR, respectively). These are regions where photons dominate the
thermal and chemical balance of the gas. Examples are O \& B stars (HII
regions), active galactic nuclei (AGN), and T Tauri
stars. In PDRs, the radiation field comprises photons with energies $6<E<13.6$
eV. Heating is provided by photo-electric emission from dust grains and cosmic
rays, while cooling proceeds through fine-structure emission lines like [OI]
63, 145 $\mu$m and [CII] 158 $\mu$m as well as emissions by H$_2$, CO and
H$_2$O rotational and vibrational lines. As a rule of thumb, a 10 eV photon
penetrates about 1/2 mag of dust.

\begin{figure}[b]
 \vspace*{-0.0 cm}
\begin{center}
 \includegraphics[width=4.2in]{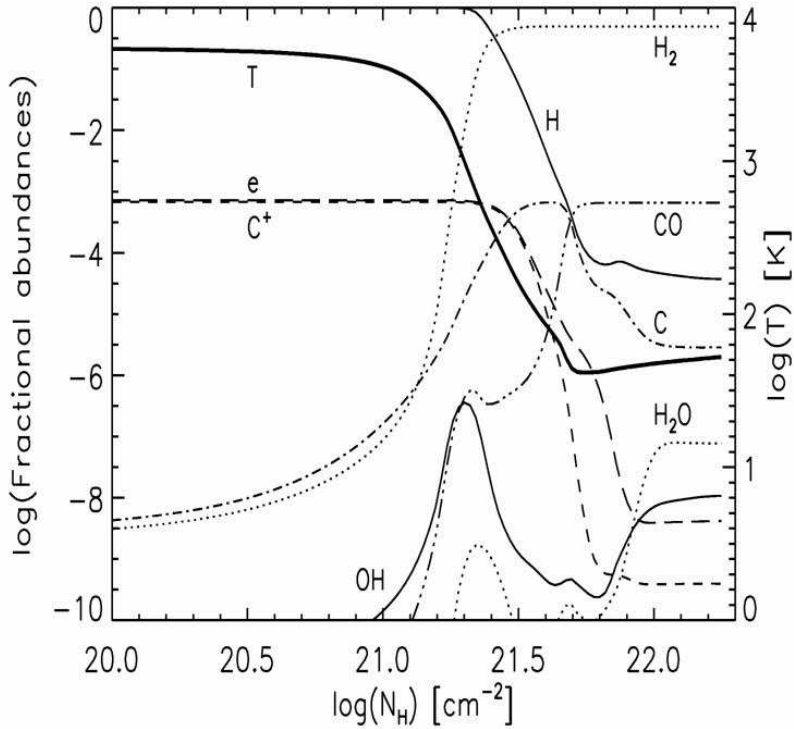}
 \vspace*{1.0 cm}
 \caption{PDR model typical of a modest starburst inside a dwarf galaxy, at a density of $10^5$ cm$^{-3}$.}
   \label{fig1}
\end{center}
\end{figure}

\begin{figure}[b]
 \vspace*{-0.0 cm}
\begin{center}
 \includegraphics[width=4.2in]{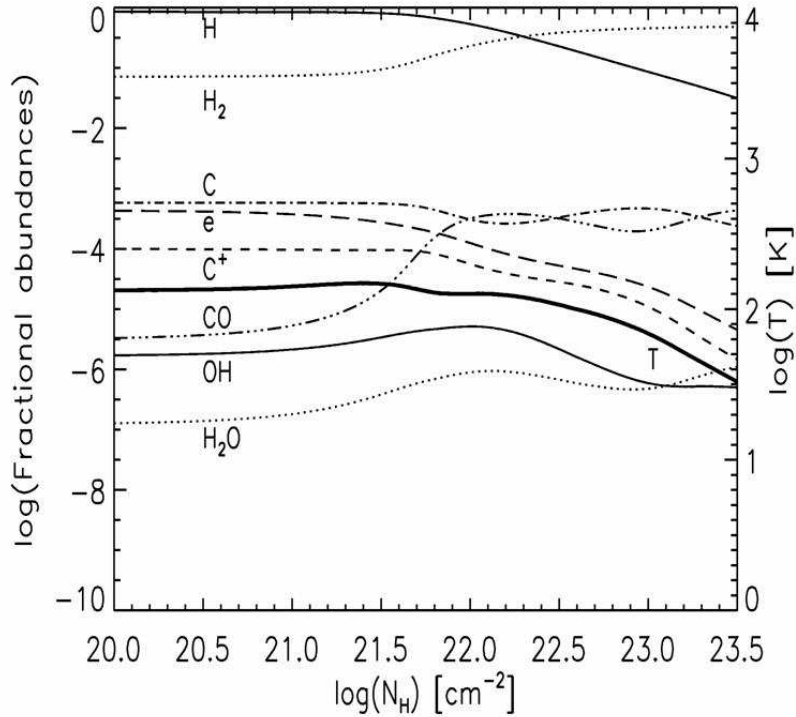}
 \vspace*{1.0 cm}
 \caption{XDR model typical of gas at a few hundred pc from a Seyfert nucleus, at a density of $10^5$ cm$^{-3}$.}
   \label{fig2}
\end{center}
\end{figure}

\begin{figure}[b]
 \vspace*{0.0 cm}
\begin{center}
 \includegraphics[width=4.2in]{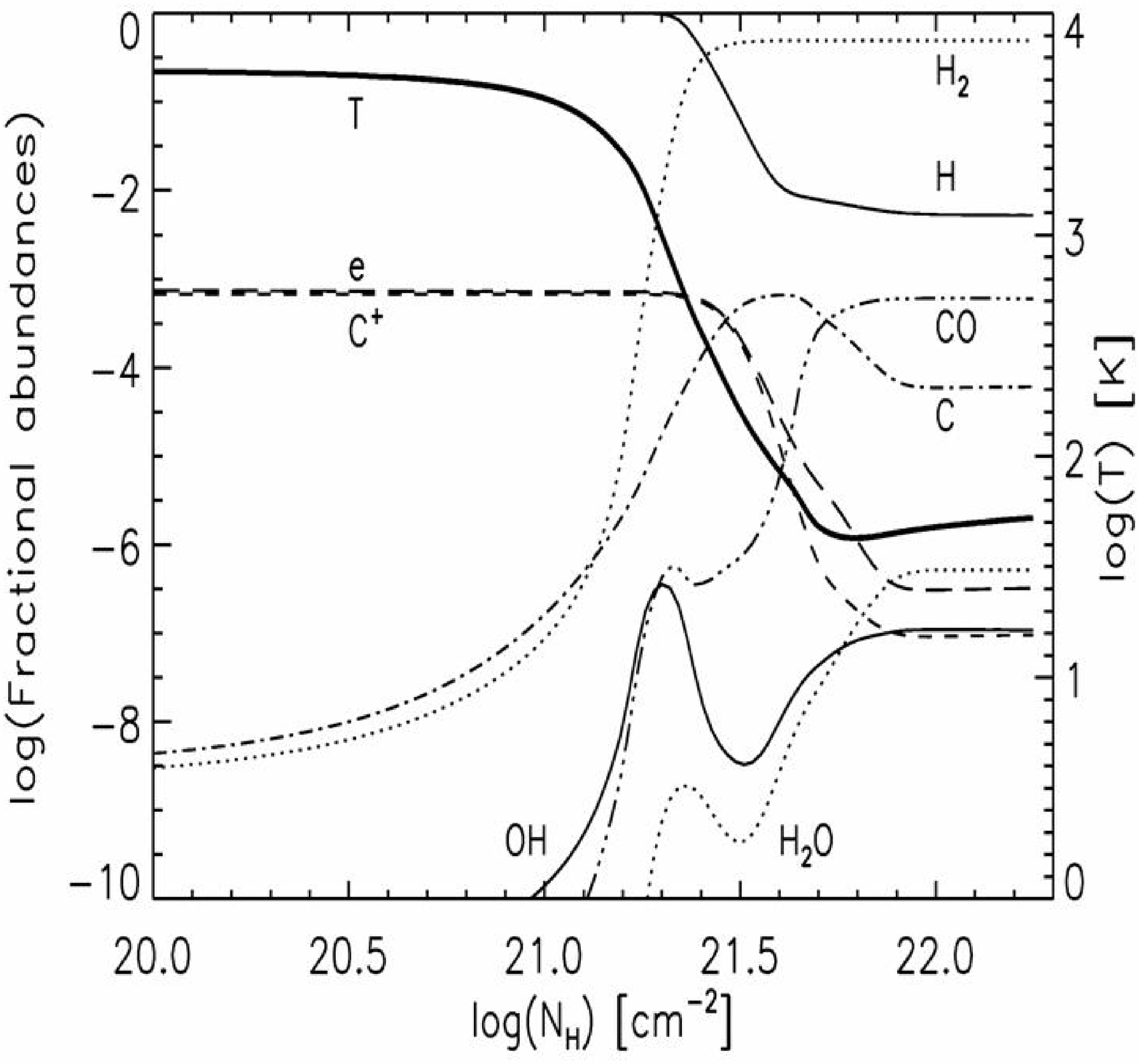}
 \vspace*{1.0 cm}
 \caption{PDR model typical of gas irradiated by cosmic rays from a supernova rate of 2 per year, at a density of $10^5$ cm$^{-3}$.}
   \label{fig3}
\end{center}
\end{figure}

In XDRs photon energies $E>0.3$ keV are considered. Heating is provided by
X-ray photo-ionizations that lead to fast electrons and Coulomb heating as
well as H and H$_2$ vibrational excitation followed by UV emission (Ly $alpha$,
Lyman-Werner); H$_3^+$ recombination heating can be important as well. Cooling
is provided by [FeII] 1.26, 1.64 $\mu$m; [OI] 63 $\mu$m; [CII] 158 and [SiII]
35 $\mu$m emission lines as well as thermal H$_2$ rotational and vibrational
emissions and gas-dust cooling. Typically, a 1 keV photon penetrates $10^{22}$
cm$^{-2}$, because cross sections scale like $~E^{-(2-3)}$.
Figures 1 and 2 show typical examples of a PDR and XDR. Note the fact that
molecules have an easier time surviving in an XDR, for the same impinging
flux by energy (Meijerink \& Spaans 2005), because molecular photo-dissociation
cross sections peak in the UV. Furthermore, the heating efficiciency in XDRs
can be 10-50\%, while it is at most 1\% in PDRs.

In figure 3 it is shown that neutral carbon is an
excellent mass tracer (as good as CO) under cosmic ray irradiations that
exceed Milky Way values by more than factor of 100 (Meijerink et al.\ 2007).
Also note that the collisional coupling between warm gas on cool dust grains
can dominate the gas cooling for modest metallicities in PDRs and XDRs.

\section{Metals}

As one lowers the metallicity one finds smaller molecular clouds and the
atomic cooling dominates by mass (Bolatto et al.\ 1999, Roellig et al.\ 2006).
The occurrence of a multi-phase medium (Wolfire et al.\ 1995) depends
strongly on metallicity and pressure (Spaans \& Norman 1997). Figure 4 shows
that only a single interstellar phase occurs for metallicities below a
percent of solar. At the same time, there is a region between 1 and 10 \%
of solar metallcity where star formation is most efficient. The reason is
that these modest metallicities allow for efficient cooling without any line
trapping (optical depth effects).

\begin{figure}[b]
 \vspace*{-0.0 cm}
\begin{center}
 \includegraphics[width=5.4in]{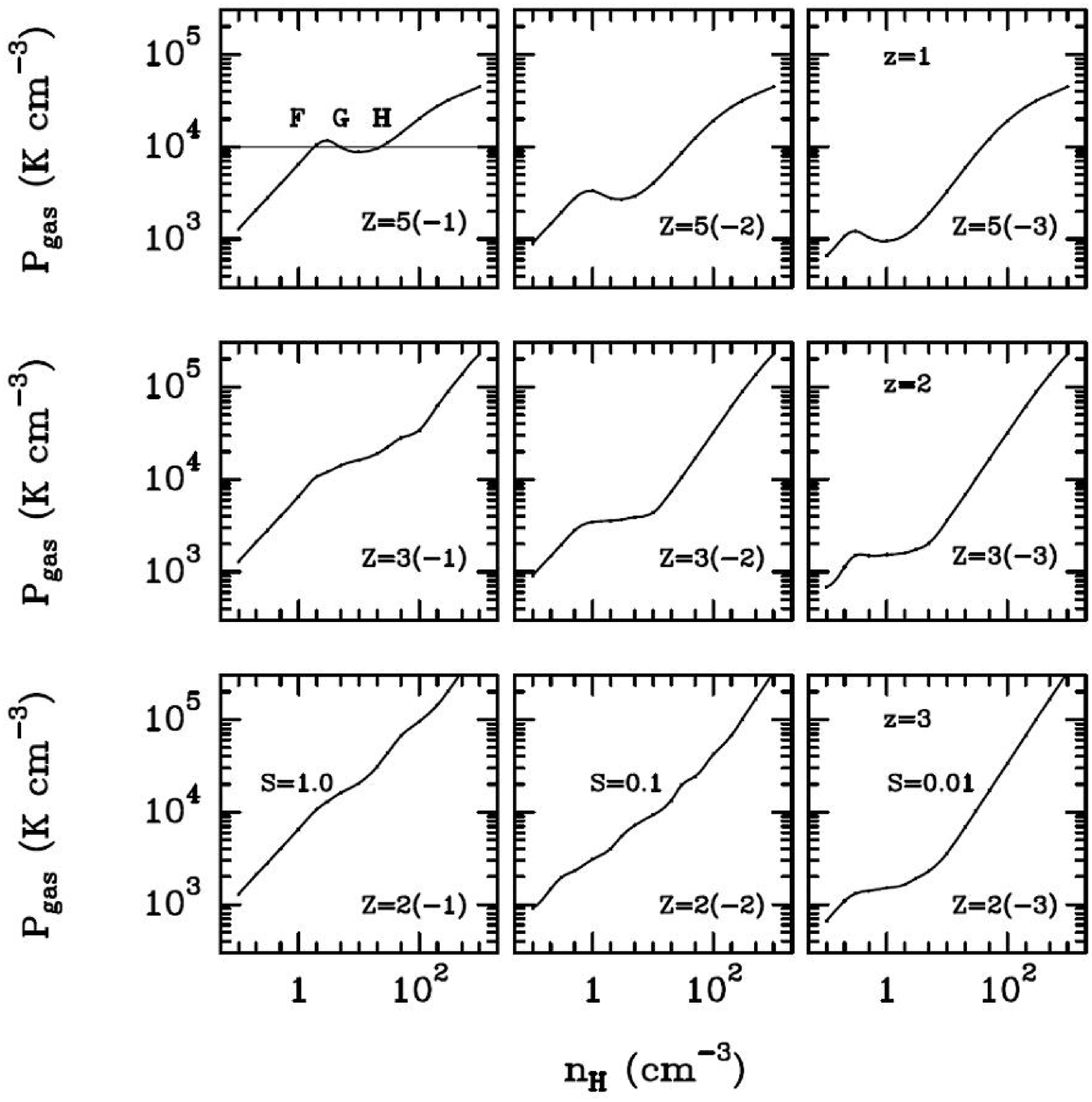}
 \vspace*{1.0 cm}
 \caption{Phase diagrams for interstellar gas as a function of metallicity and background star formation rate.}
   \label{fig4}
\end{center}
\end{figure}

\begin{figure}[b]
 \vspace*{-0.0 cm}
\begin{center}
 \includegraphics[width=6.1in]{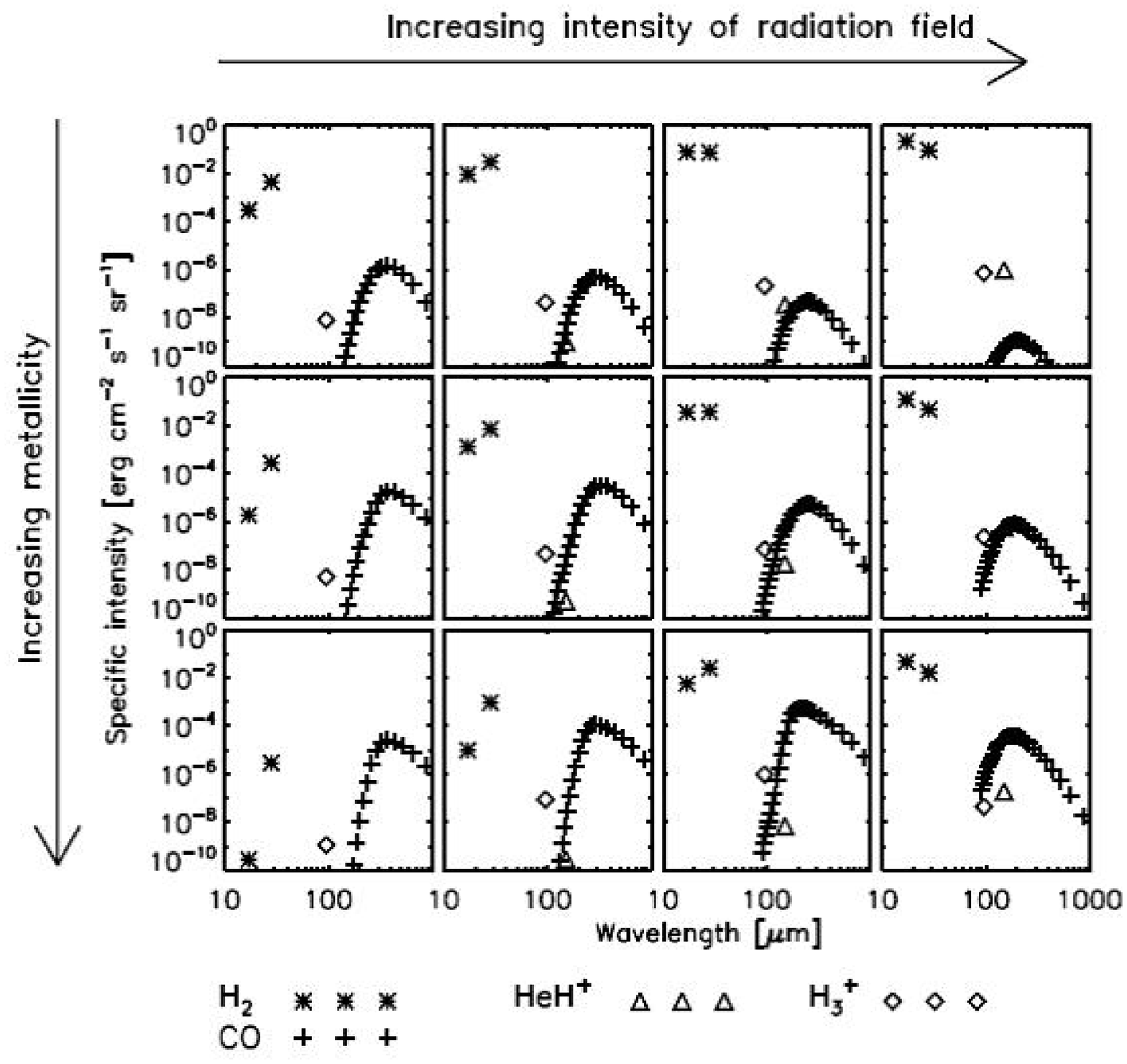}
 \vspace*{1.0 cm}
 \caption{Relative contributions from CO (metal-rich) and H$_2$ (metal-poor) gas that is irradiated by a hard spectrum of primordial galaxy; as a function of metallicity ($10^{-3}$ to $10^{-1}$/top to bottom) and impinging flux (0.1 to 100 erg cm$^{-2}$ s$^{-1}$/left to right). All panels are for a density of $10^5$ cm$^{-3}$.}
   \label{fig5}
\end{center}
\end{figure}

At metallicities well below 1\% of solar, cooling is dominated by H$_2$ and
HD emissions, which allow cooling down to $\sim 100$ K only. This is
illlustrated in figure 5, where the strengths of the first two pure rotational
H$_2$ lines are compared to the CO line spectral energy distribution, for a
system with a $10^5$ M$_\odot$ black hole accreting at Eddington. On can
clearly see that both low metallicities as well as strong irradiation favor
H$_2$ as the main coolant.

This is pertinent to the study of pop III.1 and III.2 star formation (e.g.,
Abel et al.\ 2002, Bromm et al.\ 2001; and contributions by Schneider, Ferrara
and Tan in this volume).
The relative contributions of molecular cooling depicted in figure 5 show that
the upcoming ALMA telescope will be able to see primordial systems that are
growing a massive black hole, at redshifts of $z=10-20$ (Spaans \& Meijerink
2008).

Metallicity dependent cooling is quite important for the collapse of gas
clouds and the properties of the initial mass function. in particular, LTE
effects and line trapping impact the effective equation of state (the
thermodynamics) of the gas.
This is further discussed in detail, using the FLASH code, in the contribution
of Hocuk (this volume). He finds that the level of fragmentation is a strong
function of rotational energy and metallicity.

Furthermore, metals and molecules other than H$_2$ and HD impact the formation
of structure in collapsing primordial systems. Detailed studies, using the
Enzo code, that include a complete gas-phase chemistry and formation of H$_2$
and HD on dust grains, is presented by Aykutalp (this volume). She finds that
pre-enrichment of young galaxies strongly lowers the Jeans mass of the gas
clouds they contain.

\section{Dust}

The formation of H$_2$ and HD on dust grains depends strongly on their surface
properties, as indicated in figure 6. Hydrogen atoms can be weakly bound
through van der Waals forces (physi-sorption) or strongly bound through
covalent bonds (chemi-sorption). The advantage of the latter bond is that
it allows atoms to bind to the surface even for dust temperatures well in
excess of 100 K. The hydrogen atoms either thermally hop at high dust
temperatures or tunnel at low ($\sim 10$ K) temperatures.
A comparison with the gas phase formation of H$_2$ through the
H$^-$ route, see figure 7,
indicates that dust processes dominate H$_2$ formation for
metallicities $>10^{-3.5}$ solar and electron abundances below $10^{-3}$
(Cazaux \& Spaans 2004).

\begin{figure}[b]
 \vspace*{-0.0 cm}
\begin{center}
 \includegraphics[width=5.0in]{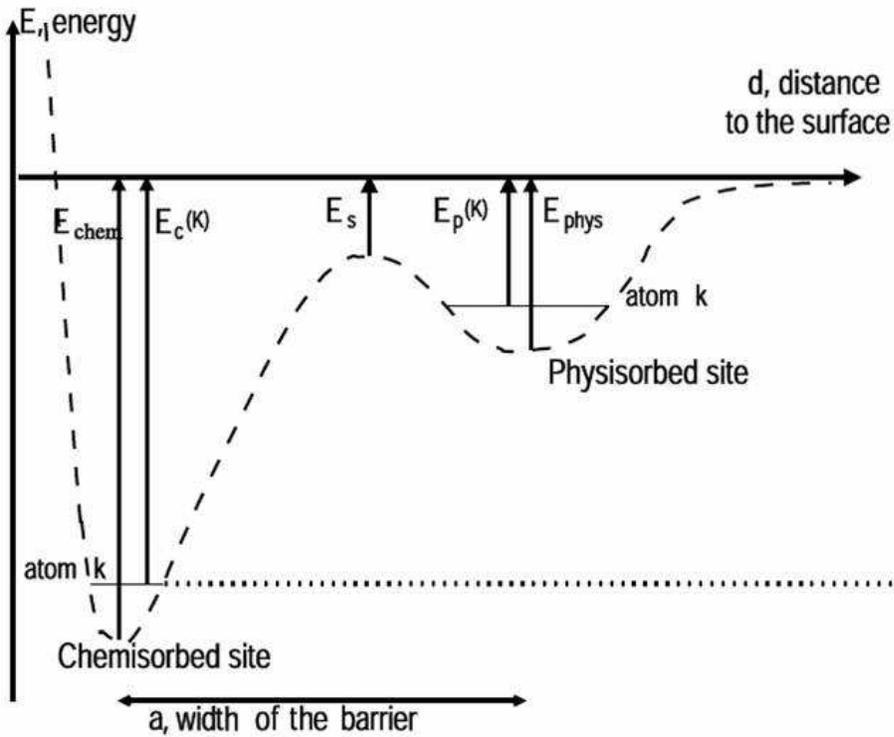}
 \vspace*{1.0 cm}
 \caption{Typical grain surface characterization (Cazaux \& Tielens 2004).}
   \label{fig6}
\end{center}
\end{figure}

The formation of HD benefits above $10^{-3}$ solar as well, as long as the
gas density is above $10^{4.5}$ cm$^{-3}$ and the electron abundance below
$10^{-3}$ (Cazaux \& Spaans 2008, in preparation). In this, the deuterium
atom is more massive that atomic hydrogen, i.e., it is less mobile and more
strongly bound to the dust grain (up to higher temperatures).

\begin{figure}[b]
 \vspace*{-0.0 cm}
\begin{center}
 \includegraphics[width=5.4in]{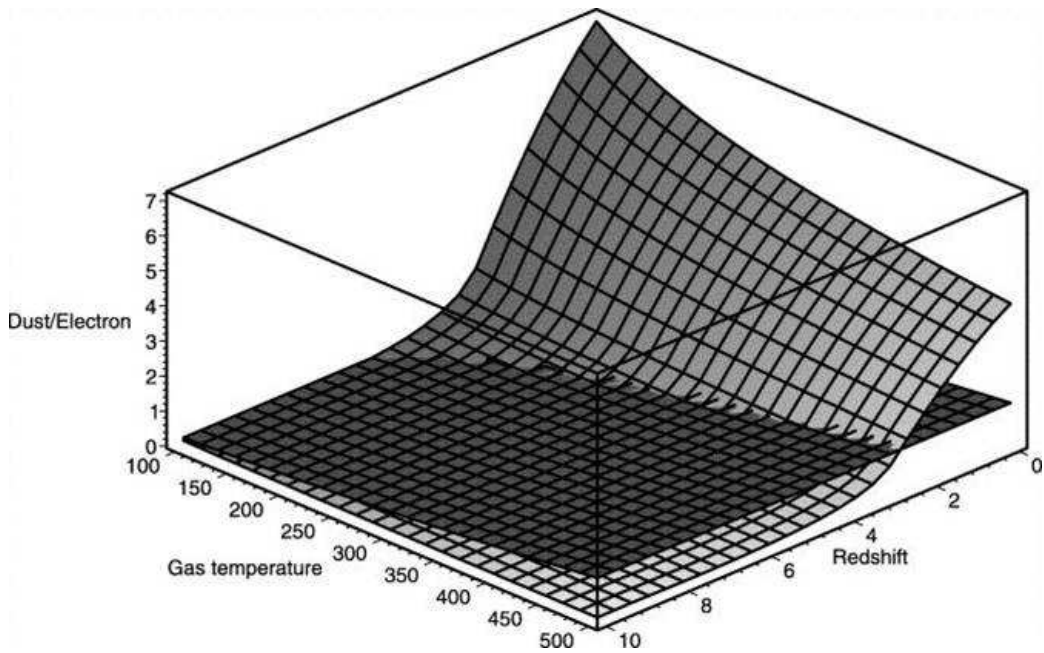}
 \vspace*{1.0 cm}
 \caption{Comparison between the gas phase and dust surface formation routes for H$_2$.}
   \label{fig7}
\end{center}
\end{figure}

\section{Conclusions}

Radiative feedback effects differ for UV and X-ray photons at any metallicity,
with molecules surviving quite well under irradiation by X-rays.
Starburst and AGN will therefore enjoy quite different cooling abilities for
their dense molecular gas. The presence of a cool molecular phase is strongly
dependent on metallicity. Strong irradiation by cosmic rays ($>200\times$ the
Milky Way value) forces a large fraction of the CO gas into neutral carbon.
Dust is important for H$_2$ and HD formation, already at metallicities of
$10^{-4}-10^{-3}$ solar.

Finally, one should always solve the equations of
statistical equilibrium to distinguish properly between the excitation,
radiation and kinetic temperature of a system. I.e., the thermodynamic floor
set by the CMB is only a hard one if the density is high enough (larger than
the critical density of a particular transition) to drive collisional
de-excitation.


\bigskip\bigskip\bigskip

\begin{thebibliography}{}

\bibitem[]{}Abel, T., Bryan, G.L., \& Norman, M.L., 2002, Science, 295, 93

\bibitem[]{}Bolatto, A.D., Jackson, J.M. \& Ingalls, J.G., 1999, ApJ, 513, 275

\bibitem[]{}Bromm, V., Ferrara, A., Coppi, P.S., \& Larson, R.B., 2001, MNRAS, 328, 969

\bibitem[]{}Cazaux, S. \& Spaans, M., 2004, ApJ, 611, 40

\bibitem[]{}Cazaux, S. \& Tielens, A.G.G.M., 2004, ApJ, 604, 222

\bibitem[]{}Meijerink, R. \& Spaans, M., 2005, A\&A, 436, 397

\bibitem[]{}Meijerink, R., Spaans, M. \& Israel, F.P., 2007, A\&A, 461, 793

\bibitem[]{}Roellig, M., Ossenkopf, V., Jeyakumar, S., Stutzki, J. \& Sternberg, A., 2006, A\&A, 451, 917

\bibitem[]{}Spaans, M. \& Meijerink, R., 2008, ApJ, 678, L5

\bibitem[]{}Spaans, M. \& Silk, J., 2000, ApJ, 538, 115

\bibitem[]{}Spaans, M. \& Norman, C.A., 1997, 483, 87

\bibitem[]{}Wolfire, M.G., Hollenbach, D., McKee, C.F., Tielens, A.G.G.M. \& Bakes, E.L.O., 443, 673

\end{thebibliography}
\end{document}